\documentclass[nofootinbib]{revtex4}
\usepackage[english]{babel}
\usepackage{array,booktabs}   
\usepackage{array} 
\usepackage{amsmath} 
\usepackage{relsize}
\usepackage{wrapfig}
\usepackage{calc}
\usepackage{sitem}
\usepackage{pdflscape}
\usepackage{color}
\usepackage{float}
\usepackage{graphicx}
\usepackage{amsmath}
\usepackage[nodisplayskipstretch]{setspace}

\usepackage{setspace}
\usepackage{tabularx}

\usepackage{float}
\usepackage{color}
\usepackage{amsmath}
\usepackage{float}
\usepackage{calc}
\usepackage{pdflscape}
\usepackage{color}
\usepackage{float}
\usepackage{graphics}
\usepackage{epsfig}
\usepackage{natbib} 

\usepackage{epstopdf}
\usepackage{amssymb}
\usepackage{graphicx}
\usepackage{epstopdf}
\usepackage{appendix}
\usepackage{romannum}
\usepackage{soul}
\usepackage{color}

\definecolor{blizzardblue}{rgb}{0.67, 0.9, 0.93}
\definecolor{bubblegum}{rgb}{0.99, 0.76, 0.8}
\usepackage[urlcolor=blizzardblue]{hyperref}
\hypersetup{
	colorlinks=false,
	linkcolor=blue,
	filecolor=magenta,      
	urlcolor=bubblegum,
}
\usepackage{relsize}

\usepackage[raggedright]{titlesec}
\begin{document}

	\pagenumbering{arabic}
	\title{Influence of cross-sectional uncertainty  on  sensitivity studies  of DUNE and T2HK experiments}
	\author{ Ritu Devi$^{1}$\footnote{Corresponding author. E-mail: rituhans4028@gmail.com}, Jaydip Singh$^{2}$, Baba Potukuchi$^{1}$}
	
	\affiliation{Department Of Physics, University of Jammu, Jammu, India$^{1}$}
	\affiliation{Department Of Physics, University of Lucknow, Lucknow, India.$^{2}$}

	\bigskip
	\begin{abstract}
		\textbf{Abstract}\\	
		The ultimate objectives of ongoing and upcoming neutrino experiments are the precise measurement of neutrino mixing parameters and the confirmation of the mass hierarchy. The systematic inaccuracy in the cross-section models introduces inaccuracy in the neutrino mixing parameters estimation. It is important to secure a large decrease of uncertainties, particularly those related to cross-section, neutrino-nucleus interactions, and  neutrino energy reconstruction, in order to achieve these ambitious goals. In this research article, we use three alternative neutrino event generators, GENIE, NuWro, and GiBUU, to analyze the sensitivity studies of T2HK, DUNE, and combined sensitivity of DUNE, and T2HK for mass hierarchy, CP violation, and octant degeneracy caused by cross-section uncertainties. The cross-section models of these generators are separate and independent. 
	\end{abstract}
	\maketitle
	
	\section {Introduction}	
	
	Neutrino oscillation physics is a well-established field of study that has faced numerous obstacles in recent decades. However, with the help of a number of remarkable experiments, significant progress has been made in the field of neutrino oscillation physics.	Neutrino oscillation is the change in neutrino flavor as they travel, i.e., a neutrino that starts off with one flavor may finish up with a different flavor after a certain distance.	Mixing angles $\theta_{ij}$, where $i<j$ = 1, 2, 3 $(\theta_{23},\theta_{12},\theta_{13})$, Dirac phase $\delta_{CP}$, and the size of mass squared difference, $\Delta m^2_{21}$ for solar mass splitting and $\Delta m^2_{31}$ for atmospheric mass splitting, are neutrino oscillation parameters that influence neutrino oscillation physics.	The near-perfect measurement of the mixing angles $\theta_{23},\theta_{12}$ and nonzero value of $\theta_{13}$ \cite{DayaBay:2012fng,DayaBay:2012yjv,RENO:2012mkc}  as well as mass squared discrepancies has made substantial progress in the exact computation of neutrino oscillation parameters. There are, however, a number of issues that are to be resolved and these are-(\romannum1) the sign of $\Delta m^2_{31}$ i.e. neutrino mass hierarchy or ordering of neutrino mass. For the neutrino mass eigenstates $m_i$ (i=1, 2, 3), there are two possible configurations. The first is Normal  Ordering (NO), also known as the Normal Hierarchy (NH), in which $m_1< m_2< m_3$, and the second is Inverted Ordering (IO), also known as the Inverted Hierarchy (IH), in which the neutrino mass order is $m_2\approx m_1>m_3$ (\romannum2) calculation of the octant of $\theta_{23}$ that refers to whether $\theta_{23}$ lies in   $0<\theta_{23}<\pi/4$ i.e. lower octant (LO) or in $\pi/4<\theta_{23}<\pi/2$ i.e. higher octant (HO). This refers to the octant degeneracy problem of $\theta_{23}$ (\romannum3) estimation of the value of the CP-violating phase $\delta_{CP}$, which can lie anywhere between $-\pi<\delta_{CP}<\pi$. As we know that CP conservation is represented by the values 0 and $\pm180^\circ$, while the maximal CP violation is represented by the value $\pm90^\circ$. This discovery may offer insight into the origins of leptogenesis \cite{Engelhard:2007kf} and may be used to answer several important puzzles, such as the universe's baryon asymmetry \cite{Pascoli:2006ie}.
	
	The requirement for precise understanding of neutrino oscillation physics depends on a variety of factors, but accurate reconstruction of neutrino energy is one of the most crucial. Any inaccurate measurement of neutrino energy will be transmitted to measurements of neutrino oscillation parameters because it introduces uncertainty into the cross section measurement and event identification. {This is because we know that the neutrino oscillation probability depends on the energy of the neutrinos.} Several notable long-baseline neutrino oscillation investigations use accelerator-produced neutrino beams. Because these neutrino	beams are not monoenergetic, complete information on the final state particles is needed to reconstruct the neutrino energy. Identifying the final state particles with the influence of nuclear effects is difficult because the particles formed at the initial neutrino-nucleon interaction vertex and the particles collected by the detector can be different or not identical. Heavy nuclear targets are used in current and anticipated future neutrino oscillation experiments to capture large event statistics, which is a prior necessity of neutrino oscillation studies, but they also amplify nuclear effects~\cite{Devi:2021udq}. The statistical error is minimized as a result of the vast event statistics obtained, and attention is directed to finding a way to handle systematic errors. One of the most common sources of systematic errors is uncertainty in the determination of neutrino-nucleus cross-sections due to the existence of nuclear effects. To decrease systematic inaccuracies, it is critical to investigate the precise neutrino-nucleon interaction cross-sections. The link between uncertainties in neutrino-nucleon cross-sections and their impact on the calculation of neutrino oscillation parameters has been studied extensively \cite{Huber:2007em,PhysRevD.87.033004,MINERvA:2004whn}.{ Several attempts have been made to study nuclear effects on neutrino oscillation measurements \cite{Fernandez-Martinez:2010erl,Benhar:2013bwa,Ankowski:2016bji,Katori:2016yel}, but in this paper we  
		investigate how three different generators for neutrino-nucleus cross-sections based on different assumptions for the nuclear dynamics affect the forecasted sensitivities to neutrino oscillation parameters at future neutrino facilities.
	}
	
	In this paper, we have chosen three alternative simulation tools, GENIE (Generates Events for Neutrino Interaction Experiments) \cite{Andreopoulos:2009rq}, NuWro \cite{PhysRevC.86.015505}, and GiBUU (Giessen Boltzmann-Uehling-Uhlenbeck) \cite{Buss:2011mx}, to represent nuclear effects and investigate the sensitivity  of the T2HK (Tokai to Hyper-Kamiokande) and DUNE (Deep Underground Neutrino Experiment) in resolving mass hierarchy, octant degeneracy, and CP violation. Nuclear effects are included in all three neutrino event generators' simulation codes, however, there are differences in the nuclear models used and the calculation of various neutrino-nucleus interaction processes. The nucleus is made up of nucleons, and it is not easy to figure out how all of them engage in neutrino-nucleus interactions. Different approximations are used to define nuclear effects by different neutrino event generators that incorporate nuclear effects in their analysis procedure.  Because the outcome of an experiment must be model-independent, this inspired us to do our analysis.
	
	The following sections comprise the paper: In Section \ref{sec2}, we discuss the NuWro, GENIE, and GiBUU neutrino event generators that were employed in this study, as well as a full comparison of their physics. In Section \ref{sec3}, we go over the simulation and experimental details, then in Section \ref{sec4}, we go over the octant sensitivity, mass hierarchy, and CP sensitivity results.
	Finally,  Section \ref{sec5}, gives our  summary and conclusions.

	\section{Simulation tools: GENIE, NuWro and GiBUU}
	\label{sec2}
\begin{figure}[htp]
		\centering
		\includegraphics[scale=0.65]{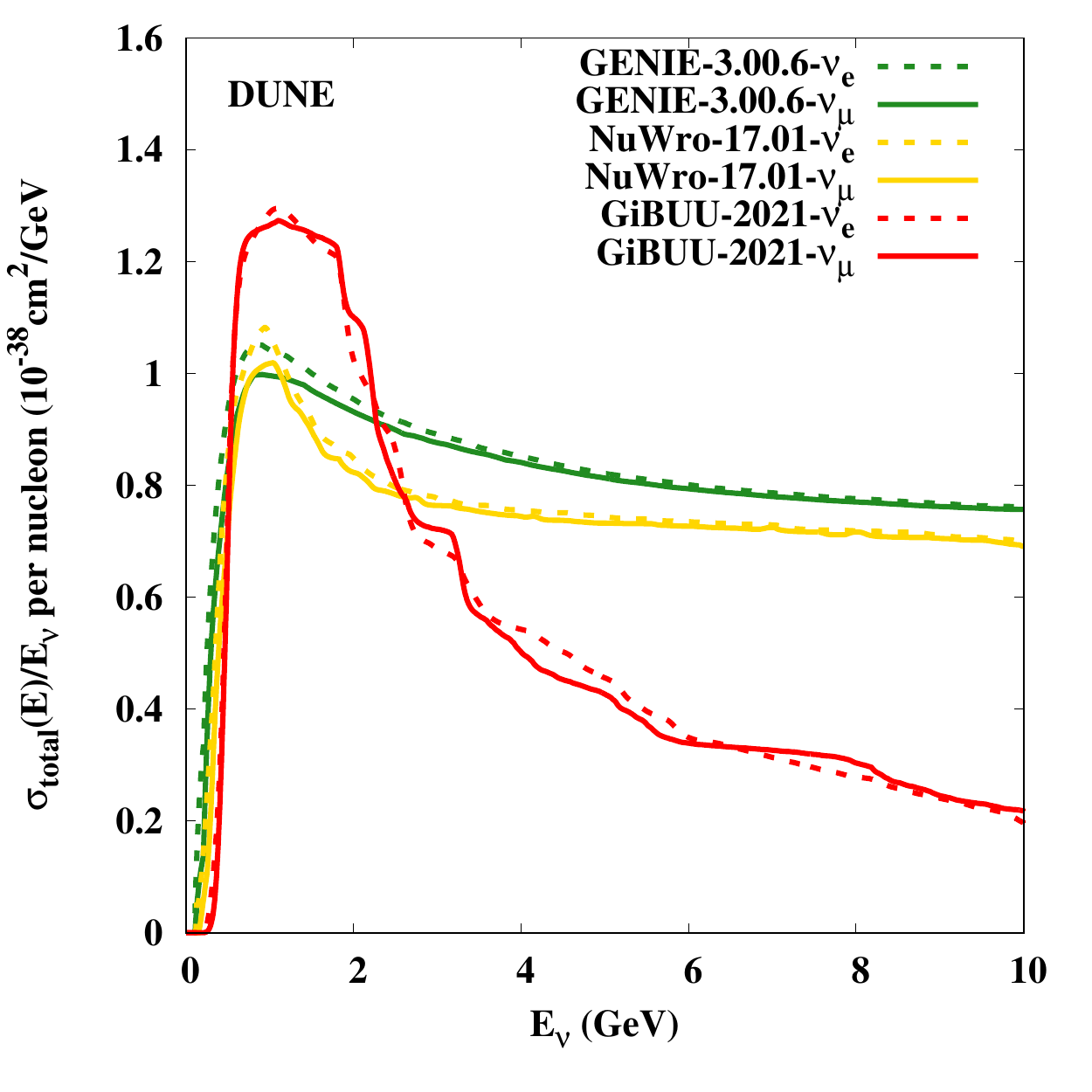}  \includegraphics[scale=0.65]{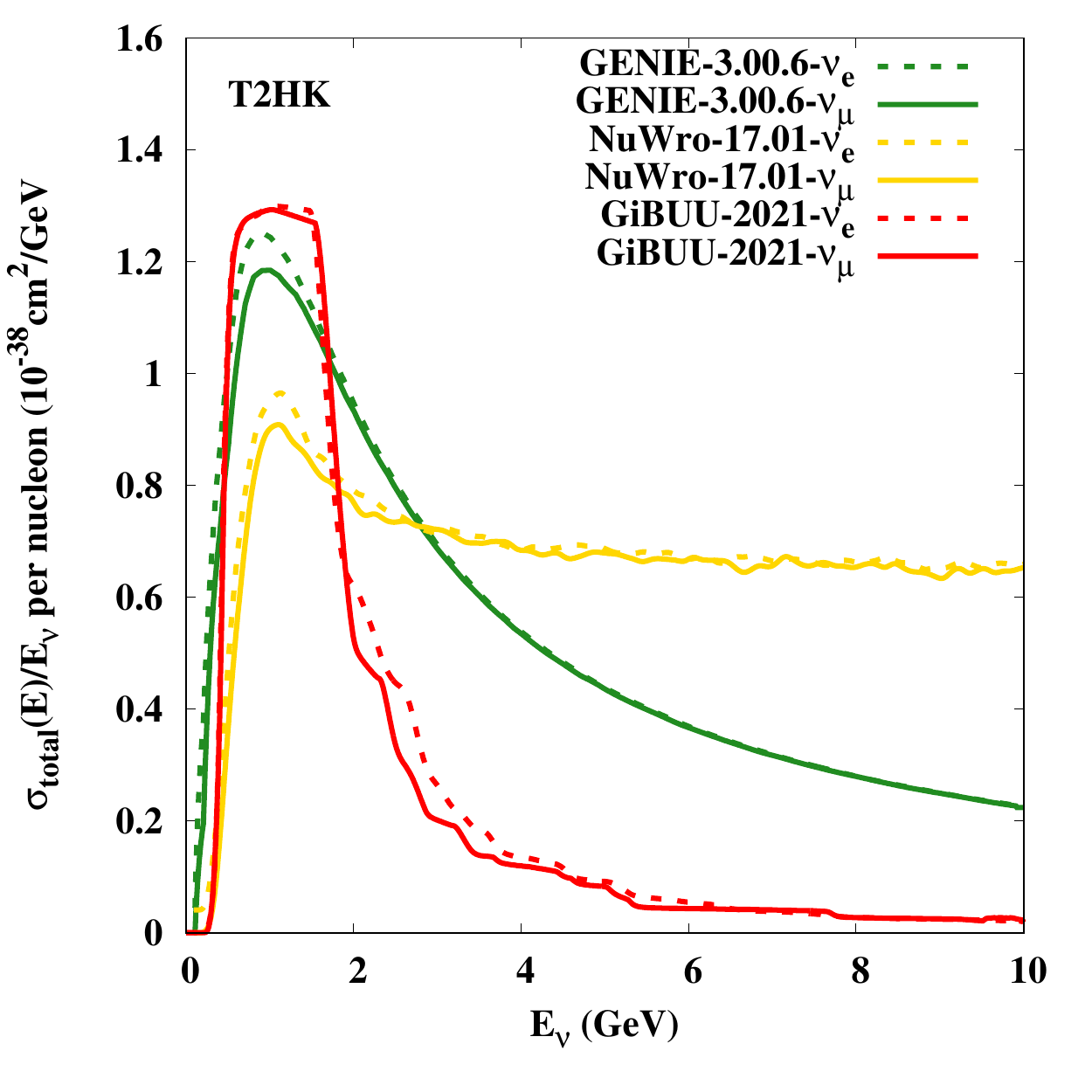} 
		\includegraphics[scale=0.65]{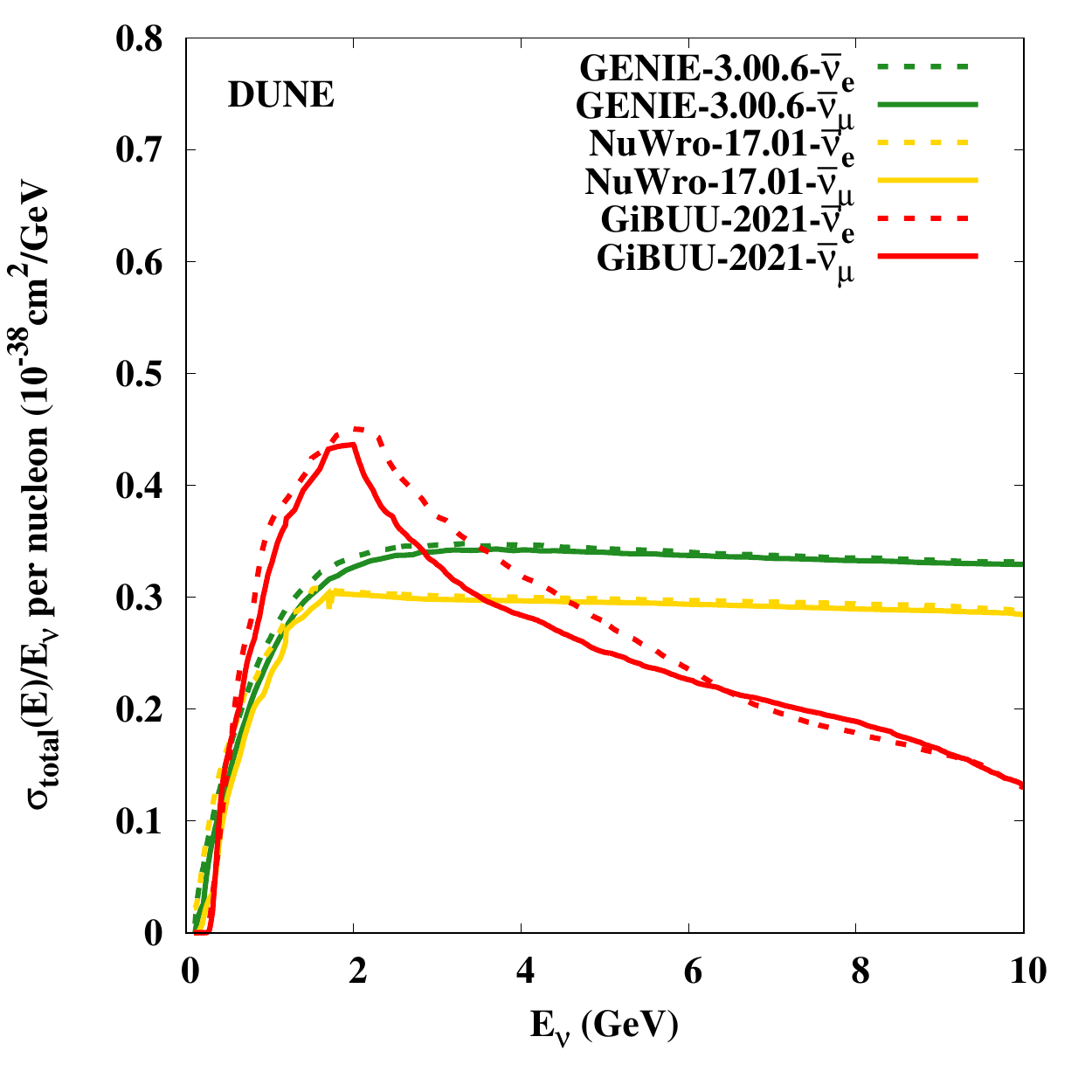}  \includegraphics[scale=0.65]{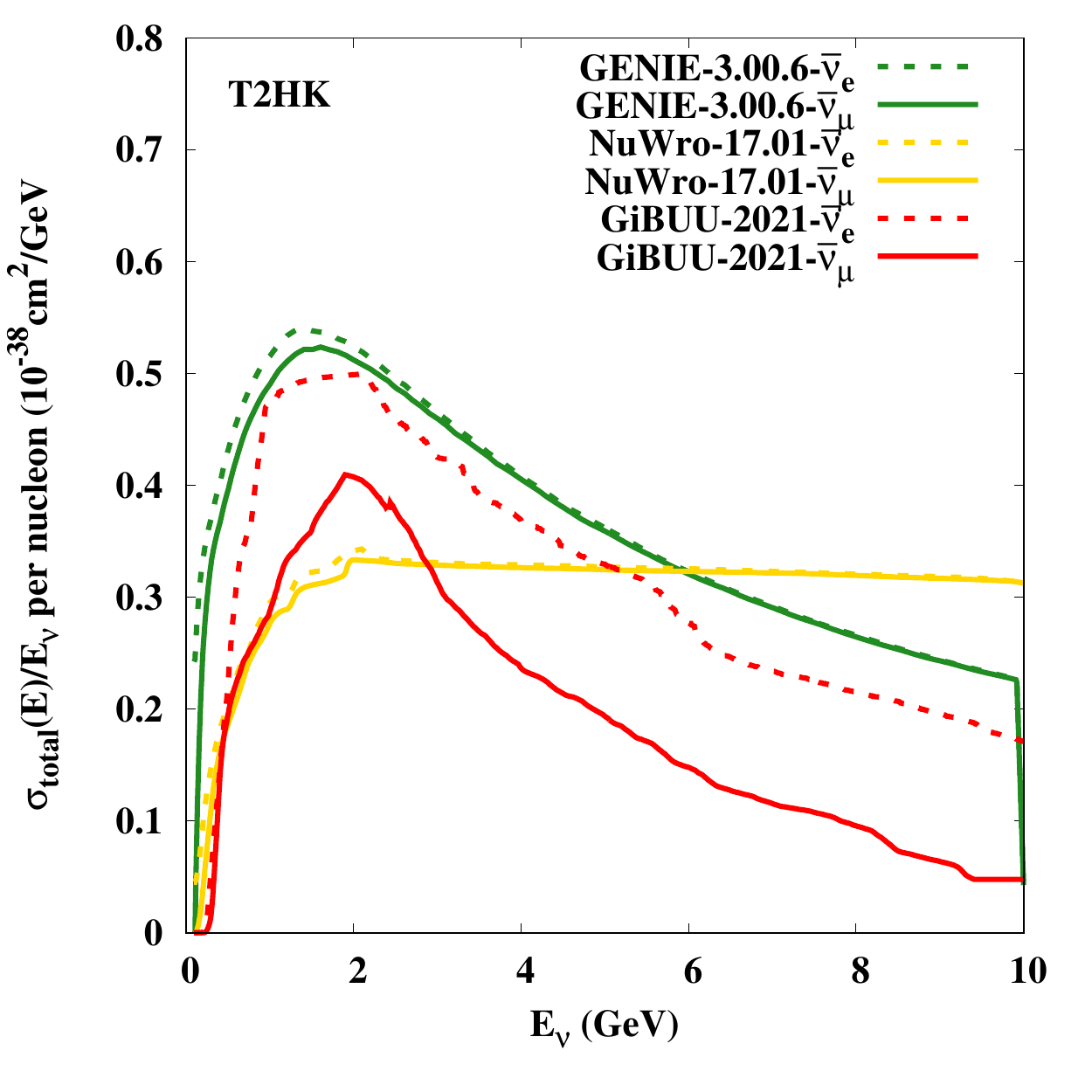}
		\caption {Total $\nu$-Ar (top left panel), $\nu$-$H_2O$ (top right panel), $\bar{\nu}$-Ar (bottom left panel) and $\bar{\nu}$-$H_2O$ (bottom right panel) interaction cross-section per nucleon as a function of true neutrino energy for GENIE (green), NuWro (yellow), and GiBUU (red).}
		\label{fig1}
	\end{figure}

	GENIE 3.00.06 \cite{Andreopoulos:2009rq}, NuWro version 19.01 \cite{PhysRevC.86.015505}, and GiBUU  v-2021 \cite{Buss:2011mx}  are the three neutrino event generators employed in this study to determine the interaction cross-section $(\nu-Ar)$ and $(\nu-H_2O)$. We looked at the interaction mechanisms of quasi-elastic (QE), resonance (RES) from resonant decay, two-particle-two-hole (2p2h/MEC), and deep inelastic scattering (DIS). The total cross-section computed from the three generators is then converted to the GLoBES (General Long Baseline Experiment Simulator) package's input format. {Figure \ref{fig1} shows the neutrino (top panel) and antineutrino (bottom panel) cross-section as a function of neutrino energy for both Ar and $H_{2}O$ targets. We see a large difference in the value of cross-sections for $\nu(\bar{\nu})-Ar$  calculated using GiBUU from GENIE and NuWro.  In the case of $\nu(\bar{\nu})-H_{2}O$, all generators show different cross-section.}
	
	In this section, we will go over the qualitative theoretical differences in nuclear models, as well as how neutrino-nucleus interaction processes are accounted for in all  generators and some popular approaches to neutrino-nucleus interaction analysis. In terms of nuclear models explaining neutrino-nucleus interactions, the selected event generators differ.

	\begin{itemize}
		\item \textbf{GENIE:} GENIE \cite{Andreopoulos:2009rq} is a neutrino event generator based on ROOT \cite{Brun:1997pa} that was written entirely in C++ utilizing object-oriented approaches. MINERvA \cite{MINERvA:2004gta}, MINOS \cite{MINOS:2007ixr}, MicroBooNE \cite{MicroBooNE:2007ivj}, NovA \cite{NOvA:2004blv}, and T2K \cite{T2K:2001wmr} are only a few of the neutrino baseline experiments that employ it around the world. For all processes, the Relativistic Fermi Gas (RFG) nuclear model is utilized. GENIE employs Bodek and Ritchie's version, which has been tweaked to include short-range nucleon-nucleon correlations \cite{PhysRevD.24.1400}. In GENIE, QE scattering is simulated by the Llewellyn-Smith model \cite{LlewellynSmith:1971uhs} with BBBA05 \cite{Bradford:2006yz}. The Rein- Sehgal model \cite{Rein:1980wg} includes the creation of baryon resonances in neutral current (NC) and charged current (CC) channels. The Feynman-Kislinger-Ravndal \cite{PhysRevD.3.2706} model of baryon resonance is used in this model. GENIE has 16 resonance modes out of 18 resonances. At neutrino energies smaller than 1 GeV, processes involving two- particle-two-hole (MEC/2p2h) excitations give an additional contribution to the overall neutrino cross-section. The nucleon-nucleon correlations in the initial state interactions (ISI), neutrino coupling to  2p2h, and  final state interaction (FSI) all play a role in this process. Simulation of  DIS processes in  GENIE is according to Bodek and Yang model \cite{Bodek:2002ps}. The axial mass in GENIE is variable, ranging from 0.99 to 1.2 $GeV/c^{2}$. 
		
		\item \textbf{NuWro:} NuWro \cite{PhysRevC.86.015505} was developed at the University of Wroclaw and has now evolved into a valuable sandbox for other generators, presenting novel theoretical models that are tested before being incorporated by NEUT and GENIE. For the QE process, NuWro uses fundamental Llewellyn Smith formulas \cite{LlewellynSmith:1971uhs}, with a wide range of vector form factor possibilities (dipole, BBA03 \cite{Budd:2003wb}, BBBA05 \cite{Bradford:2006yz}, Alberico et al. \cite{PhysRevC.79.065204}). Typically, global and local RFG models or the spectral function (SF) method are utilized. With nucleon-$ \Delta $ form factors for RES events given in Ref. \cite{PhysRevD.80.093001}, only the $ \Delta $ resonance is explicitly provided, with $ C^{5}_{A}(0) $=1.19 as the free parameter and for the rest of the resonance modes, we have an average based
		on the Adler-Rarita-Schwinger model \cite{PhysRevD.80.093001} for RES events. For simulation of the DIS channel$(W>1.6GeV)$, the Bodek-Yang prescription \cite{Bodek:2002vp} is used to analyze the total cross-sections. For specified quark configurations, the Pythia6 hadronization routine is used to allow their employment in the low W area down to 1.2 GeV.
		
		\item \textbf{GiBUU:} With a single, consistent physics model, GiBUU \cite{Buss:2011mx} intends to represent a large number of different nuclear processes (electron, proton, photon, pions, neutrino, A)-A  over a wide range of energies. It employs FORTRAN procedures. Ref. \cite{Buss:2011mx} contains the most detailed description of the model.  The RFG model in GiBUU is updated by including a density-dependent mean-field potential term that assumes all nucleons are bound. Ref.\cite{ PhysRevC.86.054606,PhysRevC.94.035502} provides more information on the QE cross-section process in GiBUU. $ M_A=1 GeV/c^2  $is the axial mass employed by GiBUU.
		The impulse approximation is used to model true CCQE (charged current quasi-elastic) neutrino interactions with single nucleons. For the axial component, a typical dipole form is utilized, while the BBBA07 \cite{Bodek:2007wz} parameterization is used for the vector form factors. Based on the kinematics of the final state nucleons, the computation also employs a density-dependent Pauli-blocking of interactions.
		There are 13 different types of resonance modes in GiBUU. MAID analysis of electron scattering data \cite{Drechsel:1992pn,Rein:1981ys} yields vector form factors for each of GiBUU's 13 resonance modes. Simulation of DIS process in GiBUU is done by PYTHIA6 \cite{Lalakulich:2011eh}.
		
	\end{itemize}

	\section{Experimental details for the simulation studies}
	\label{sec3}

	We simulated the tests alone and  combined them to better understand the sensitivities and complementarity of DUNE and T2HK. We are using the GLoBES libraries \cite{Huber:2004gg,Huber:2009cw}, which require cross-section, neutrino, and anti-neutrino beam fluxes, and detector parameterization parameters as input. The cross-section input format is  $\hat{\sigma}(E) = \sigma(E)/E [10^{-38}{cm^{2}}/{GeV}]$, and more information can be found in \cite{globes}. The experimental specifications for the  DUNE and T2HK setups that we employed in our analysis are listed below.
	
	\begin{itemize}
		\item \textbf{T2HK}: The  T2HK  experiment \cite{Hyper-Kamiokande:2016dsw} is a proposed next-generation long-baseline experiment that will use a neutrino beam generated at J-PARC in Tokai and directed 2.5 degrees off-axis to Hyper-Kamiokande (Hyper-K). The target power is supposed to be 4 MW, with a running period of 5 years for an antineutrino beam and 5 years for a neutrino beam. The planned HyperKamiokande detector, with a fiducial mass of 500 kton  and a baseline length of 295 km, is used as a detector. The narrow-band beam is largely made up of $\nu_{\mu}$ (or $\bar{\nu}_{\mu}$), with a peak energy of 0.6 GeV, which corresponds to the first oscillation maximum at 295 km. At this energy range, QE is the most prevalent process which itself has systematics from 2p2h/MEC channels~\cite{Singh:2019qac}.  For simulation, our input files for T2HK are based on the GLoBES package, which has been extensively modified to meet the most recent experimental design. In both signal and background normalization, we account for 5\% uncertainty {for both the appearance and disappearance channels} \cite{PhysRevD.72.033003}. The energy resolution for both $ \nu_{e} $ and $ \nu_{\mu} $ is 8.5\%/$\sqrt{E}$ GeV \cite{Huber:2002mx}.
		
		\item \textbf{DUNE:} DUNE at LBNF (Long-Baseline Neutrino Facility) will be made up of a near detector (ND) and far detector (FD) with the same  Argon (Ar) material, but the proportions and technology will change.
		The ND system will be positioned 574 meters downstream and 60 meters underground \cite{DUNE:2020jqi} from the neutrino beam source site at Fermilab.  We assumed a FD with a fiducial volume of 40 kton liquid argon (A = 40) located at a distance of L = 1300 km from the wide-band neutrino beam source and an operating time of 5 years in both neutrino and antineutrino mode for the  simulation of DUNE experiment. The DUNE-LBNF flux has an average energy of 2.5 GeV and the RES process is the most prevalent process at this energy. The neutrino flux employed in this study corresponds to the 80 GeV Reference beam configuration \cite{cdr}, with a beam power of 1.07 MW. An energy smearing technique, which we decided to be a Gaussian function of energy resolution, is used to compute the binned event rates \cite{globes}. The energy resolution for $\nu_{e}$ is 15\%/$\sqrt{E}$ (GeV), while for $\nu_{\mu}$ it is 20\%/$\sqrt{E}$  (GeV) \cite{LBNE:2013dhi}.We account for 5\% uncertainty in  the signal and 10\% in  background normalization { for both the appearance and disappearance channels.} Table \ref{table1} shows the true values of the oscillation parameters \cite{Esteban:2018azc} considered in this study.

	\end{itemize}
	
	\begin{table}[htp]
		\caption{Oscillation parameters included in our analysis.}
		\renewcommand\thetable{\Roman{table}}
		\centering
		\setlength{\tabcolsep}{2pt}
		\begin{tabular}{c | c | c }
			
			\hline\hline
			\textbf{Parameter}           &  \textbf{True value}      & \textbf{Marginalization Range}\\
			\hline
			$\theta_{12}$               &           0.590                & Not marginalized                             \\
			$\theta_{13}$               &           0.151                & Not marginalized                            \\
			$\theta_{23}$(NH)           &           0.867                &  [0.703, 0.914]  \\
			$\theta_{23}$(IH)           &           0.870                &  [0.710, 0.917]  \\
			$\delta_{CP}$               &           0                    &  [-$\pi, \pi$] \\
			$\Delta m^{2}_{21}$         &           7.39 $\times$ $10^{-5}$  $eV^{2}$      & Not marginalized                            \\
			$\Delta m^{2}_{31}$(NH)     &           2.525 $\times$ $10^{-3}$  $eV^{2}$    & [2.427, 2.625] $\times$ $10^{-3}$  $eV^{2}$ (NH)    \\
			$\Delta m^{2}_{31}$(IH)     &          -2.512 $\times$ $10^{-3}$  $eV^{2}$   & -[2.611, 2.412] $\times$ $10^{-3}$  $eV^{2}$ (IH)   \\
			\hline\hline
		\end{tabular}
		\label{table1}
	\end{table}

	\section{Sensitivity studies}
	\label{sec4}
	
	In this section, we attempt to investigate the impact of cross-sectional uncertainties (deficiencies in the theoretical aspects of nuclear physics models as implied by the generators) on the three key goals of DUNE and T2HK, namely (i)  CP phase violation, (ii) mass ordering, and (iii) octant degeneracy. Some of these aspects have already been studied in detail by various authors \cite{PhysRevD.76.031301,PhysRevD.87.053006,Barger:2013rha,Ballett:2016daj,Deepthi:2014iya,CHAKRABORTY2018303,NAGU2020114888}. We estimate the sensitivity of each experiment in terms of $\chi^2$. { We calculate the statistical $\chi^2$ by comparing the true events $N^{true}$ and test events $N^{test}$ using the following Poisson formula:
		\begin{equation}
		{\chi}_{stat}^2 = \sum_i 2 \left[N_i^{test}- N_i^{true} - N_i^{true} log \left ( \frac{N_i^{test}}{N_i^{true}}\right) \right]
		\end{equation}
		where the index $i$ corresponds to the number of energy bins.
		To incorporate the effect of the systematics, we deviate the
		test events by
		\begin{equation}
		N_i^{test} \rightarrow N_i^{test} \left(1+\sum_k c_i^k \xi_k\right)
		\end{equation}
		where $c_i^k$ is the 1$\sigma$ systematic error corresponding to the pull variable $\xi_k$. Here the index $k$ stands for number of pull variables. After modifying the events the combined statistical and systematic $\chi^2$ is calculated as
		\begin{equation}
		\chi_{stat+sys}^2 = \chi_{stat}^2+ \sum \xi_k^2
		\end{equation}
		The final $\chi^2$ is obtained by varying $\xi_k$ from -3 to +3 corresponding to their 3$\sigma$ ranges and minimizing over $\xi_k$  i.e.,
		\begin{equation}
		\chi^2 = min\{\xi_{k}\}\left[\chi_{stat+sys}^2\right]
		\end{equation}
		For our analysis of the both experiments we take four
		pull variables. These variables are: (i) signal normalization error, (ii) signal tilt error, (iii) background normalization error and (iv) background tilt error. We do not use any prior on any parameters.}
	
	\begin{figure}[htb!]
		\centering
		\includegraphics[scale=0.7]{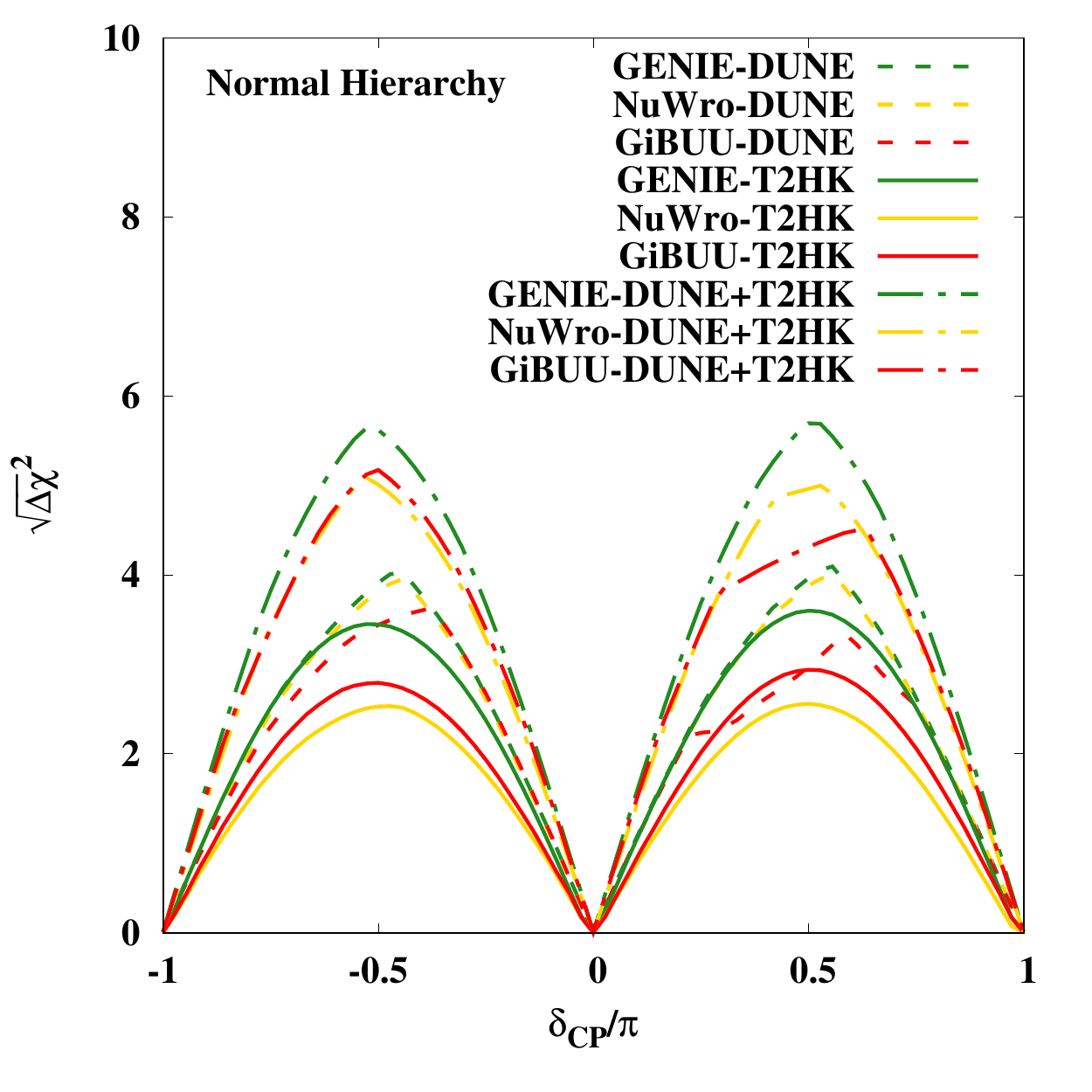} 	\includegraphics[scale=0.7]{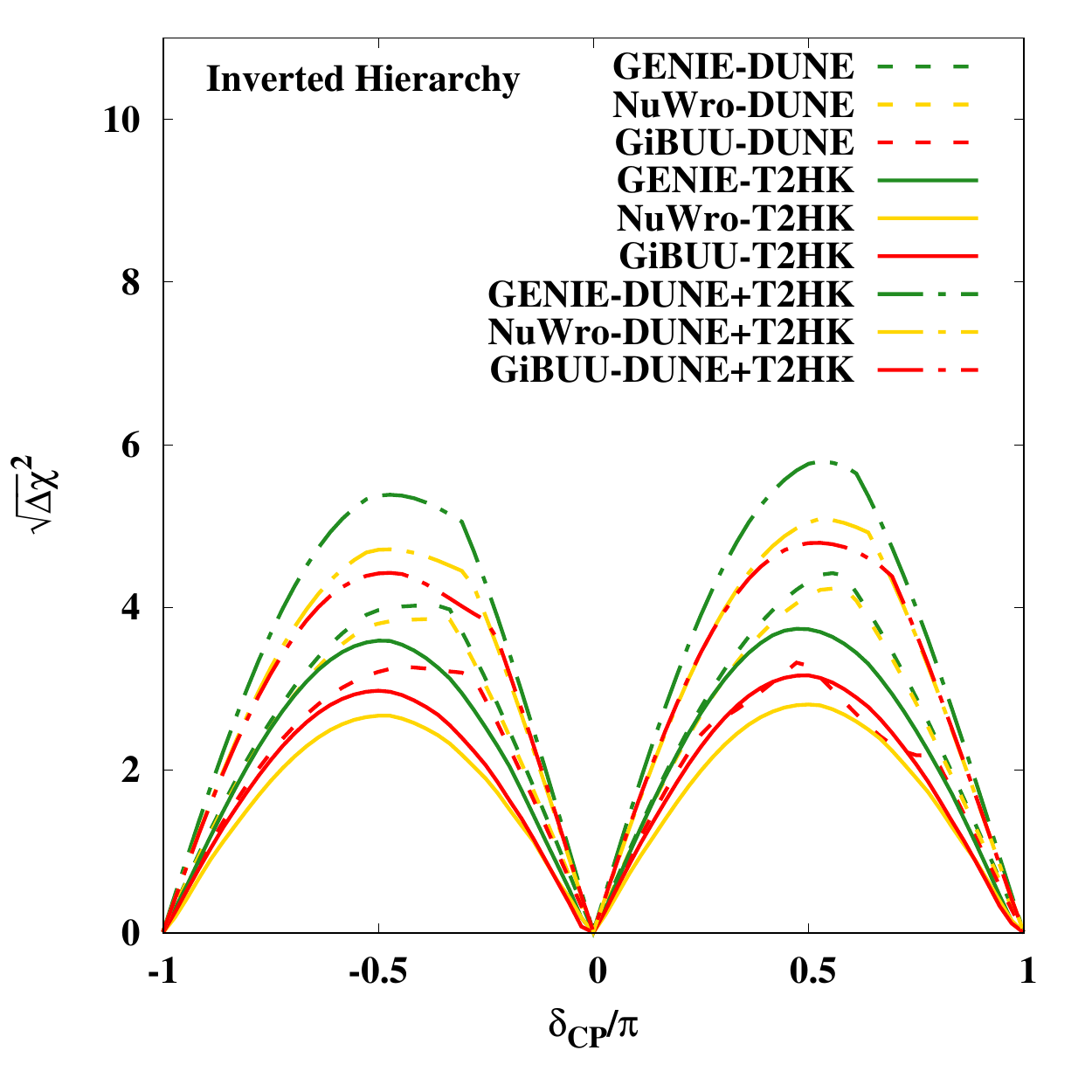} 
		\caption {CP sensitivity measurement as a function of the true value of $ \delta_{CP} $ for NH (left panel) and IH (right panel) by GENIE (green lines), NuWro(yellow lines), and GiBUU (red lines) for T2HK, DUNE, and T2HK+DUNE experiments. }
		\label{fig2}
	\end{figure}

	\subsection{CP violation sensitivity with DUNE and T2HK}

	To detect CP violation, the CP phase value must be different from CP preserving values, such as 0 or $\pm \pi$. Because the true value of $ \delta_{CP} $ is unknown, the analysis is carried out by scanning all possible true values of $ \delta_{CP} $ over the complete range $-\pi<\delta_{CP}<+\pi$ and comparing them to the $ \delta_{CP} $ conserved values. $ \delta_{CP} $, $\theta_{23}$ and $|\Delta m^2_{31} |$ are our test parameters. We minimized over the test parameters in the 3$ \sigma $ range when doing our analysis, as shown in Table \ref{table1}. We do the following computations to get the CP violation sensitivity:
	
	\begin{equation} \label{eq1}
	\Delta \chi^{2}_{0} = \chi^{2}(\delta_{CP}=0) - \chi^2_{true}
	\end{equation}
	\begin{equation}\label{eq2}
	\Delta \chi^{2}_{\pi} = \chi^{2}(\delta_{CP}=\pi) - \chi^2_{true} 
	\end{equation}
	\begin{equation}\label{eq3}
	\Delta \chi^{2} = min(\Delta \chi^{2}_{0},\Delta \chi^{2}_{\pi})
	\end{equation}
	
	Using $\sigma=\sqrt{\Delta\chi^2}$, a qualitative handle on the measurement of CP violation is acquired, as shown in the Fig. \ref{fig2}.  
	
	{ The CP sensitivity  for normal hierarchy (left panel) and inverted hierarchy (right panel) is shown in  Fig.\ref{fig2}.  In the DUNE experiment, GiBUU shows only 1$\sigma$ difference from GENIE and NUWro, whereas in case of T2HK experiment,  a  difference of around 1$\sigma$ between GENIE and NuWro at $\delta_{CP} \sim \pm$0.5/$\pi$  for both the NH and IH cases. When we combine DUNE and T2HK, we see a small difference in all generators for the NH but for the IH, GENIE shows 1$\sigma$ difference  from  GiBUU and NuWro at $\delta_{CP} \sim \pm$0.5/$\pi$.
	}
	
	\subsection{Mass hierarchy sensitivity}
\begin{figure}[htp!] 
		\centering
		\includegraphics[scale=0.7]{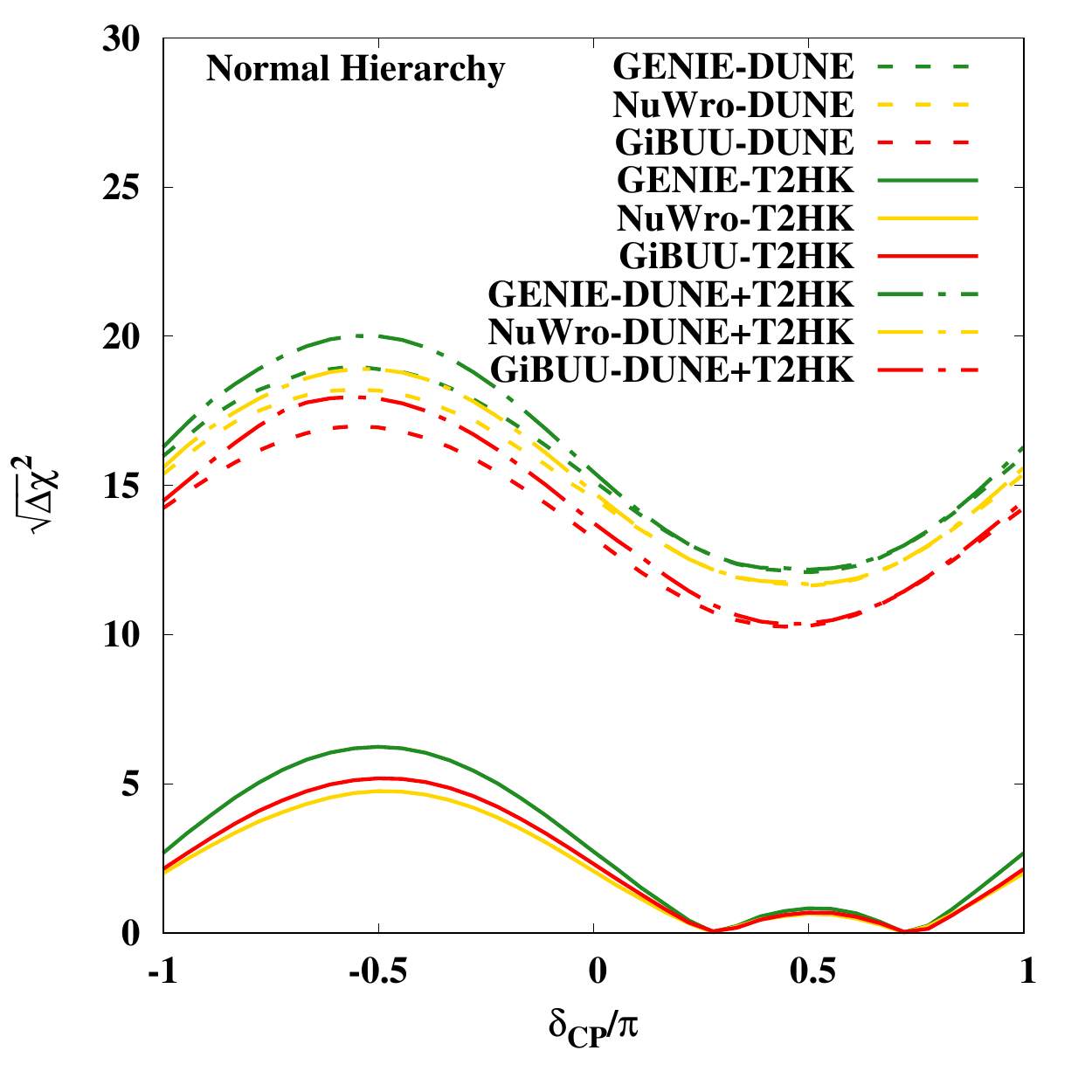} 	\includegraphics[scale=0.7]{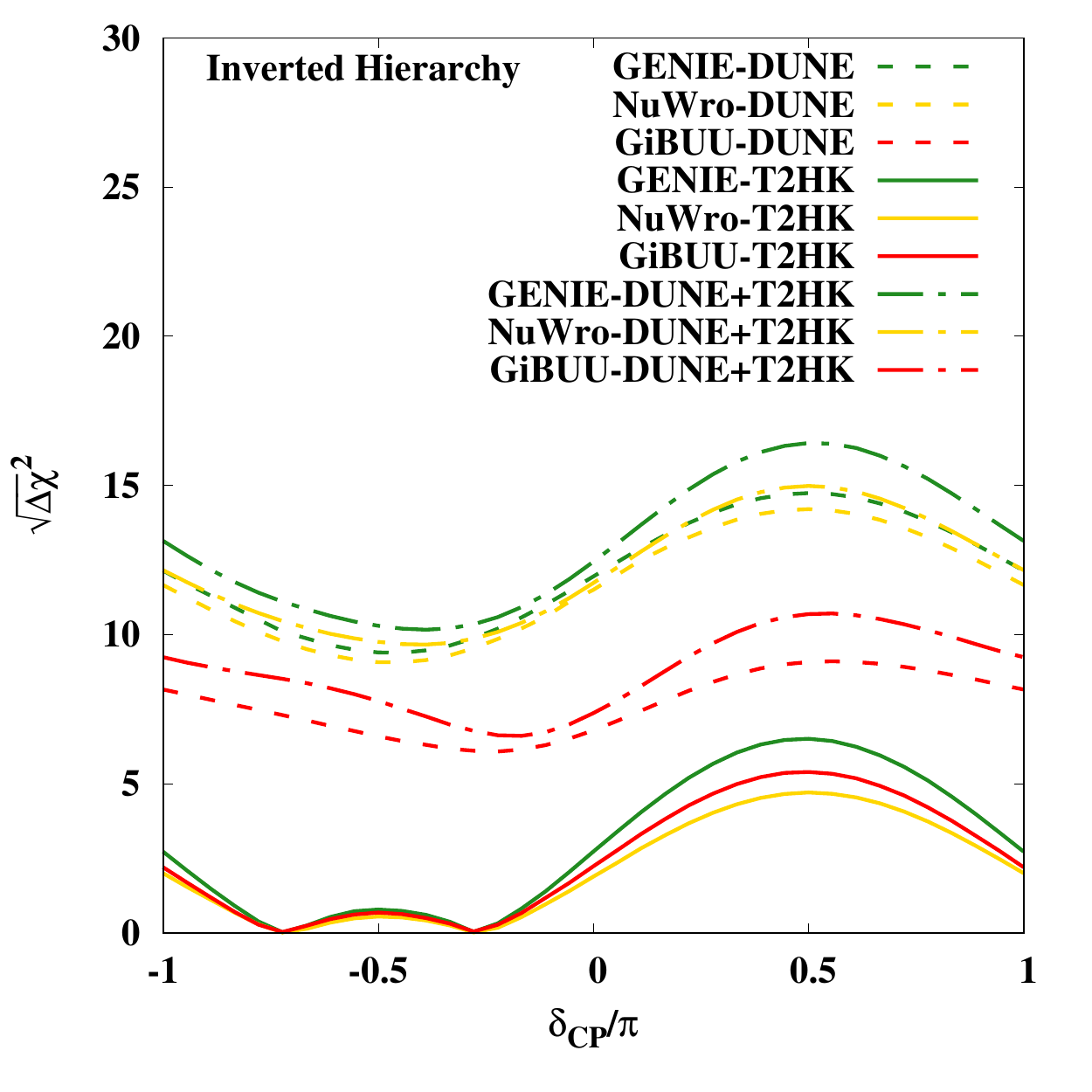} 
		\caption {Mass hierarchy sensitivity measurement as a function of the true value of $\delta_{CP}$ for NH (left panel) and IH (right panel).}
		\label{fig3}
	\end{figure}

	The quest to understand the real nature of neutrino mass ordering, whether normal or inverted, is one of the most important topics in neutrino physics. The sensitivity of the  mass hierarchy is evaluated by assuming a normal (inverted) hierarchy to be a true hierarchy and comparing it to an inverted (normal) hierarchy using equations \eqref{eq4}, \eqref{eq5}. As a result, we create hierarchies with true and test values that are diametrically opposed. The mass hierarchy sensitivity is shown in Fig. \ref{fig3} for both the normal (left panel) and inverted (right panel) cases. The following formula is used to compute the $ \Delta \chi^{2} $ quantity for mass hierarchy sensitivity:  
	
	\begin{equation} \label{eq4}
	\Delta\chi^{2}_{MH} = \chi^{2}_{IH} - \chi^{2}_{NH} 
	\end{equation}
	\begin{equation} \label{eq5}
	\Delta\chi^{2}_{MH} = \chi^{2}_{NH} - \chi^{2}_{IH} 
	\end{equation}

	{From figure \ref{fig3}, we notice that  in the case of the T2HK experiment, all generators show similar results for both mass hierarchy cases.
		In the case of the DUNE experiment, there is around 2$\sigma$ difference for the NH but there is around  6$\sigma$ difference between GENIE and GiBUU for the IH case.  In the combined DUNE and T2HK result, we see a 6$\sigma$ difference between GENIE and GiBUU for the IH case, but only a 2$\sigma$ difference for the NH case.}
	
	\subsection{Octant sensitivity}
	
	The atmospheric mixing angle $ \theta_{23} $ is not yet proven to be in the lower octant $ (0<\theta_{23}< \pi/4) $-LO or the higher octant $ (\pi/4<\theta_{23}<\pi/2 ) $-HO, with a {maximal} value of $\pi/4$.
	The development of multiple disconnected zones in the multidimensional neutrino oscillation parameter space is the primary challenge in resolving the octant degeneracy. Parameter degeneracy occurs when it is impossible to locate the exact or real answer for a given collection of true values. The test value of $ \theta_{23} $ is  modified in the lower (higher) octant range while doing the octant sensitivity calculations in the lower (higher) octant range. The true value of $ \theta_{23} $ is 0.703 in the lower octant and 0.867 in the higher octant, although the range of test values for $ \theta_{23} $ in LO and HO is [0.785, 0.961] and [0.609, 0.785], respectively. The metric $\Delta \chi^2$ for octant sensitivity is defined as
	
	\begin{equation}
	\Delta\chi^{2}_{octant} = |\chi^{2}_{\theta_{23}^{test}>\pi/4} - \chi^{2}_{\theta_{23}^{true}<\pi/4}|
	\end{equation}
	
	\begin{figure}[htp]
		\centering
		\includegraphics[scale=0.7]{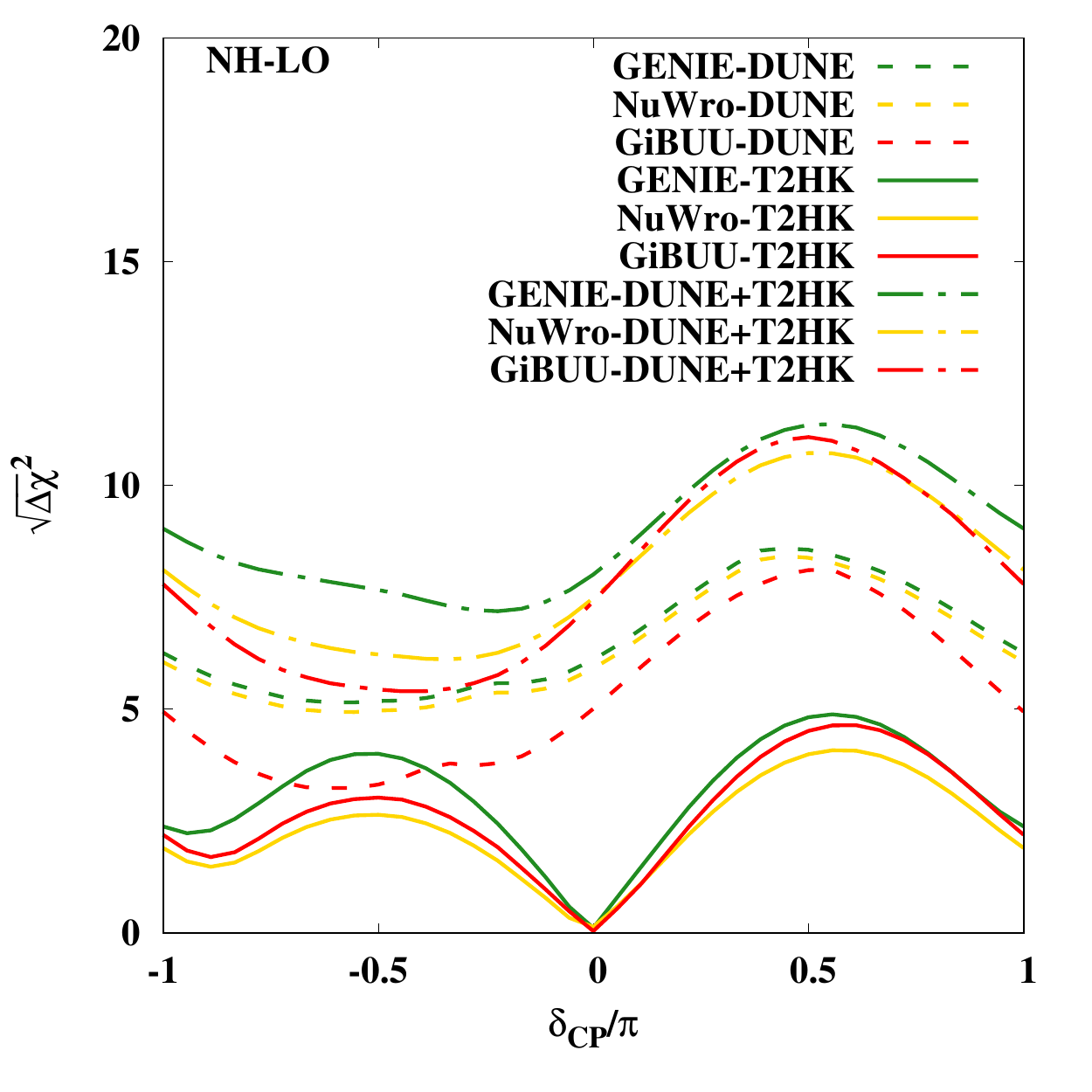} 	\includegraphics[scale=0.7]{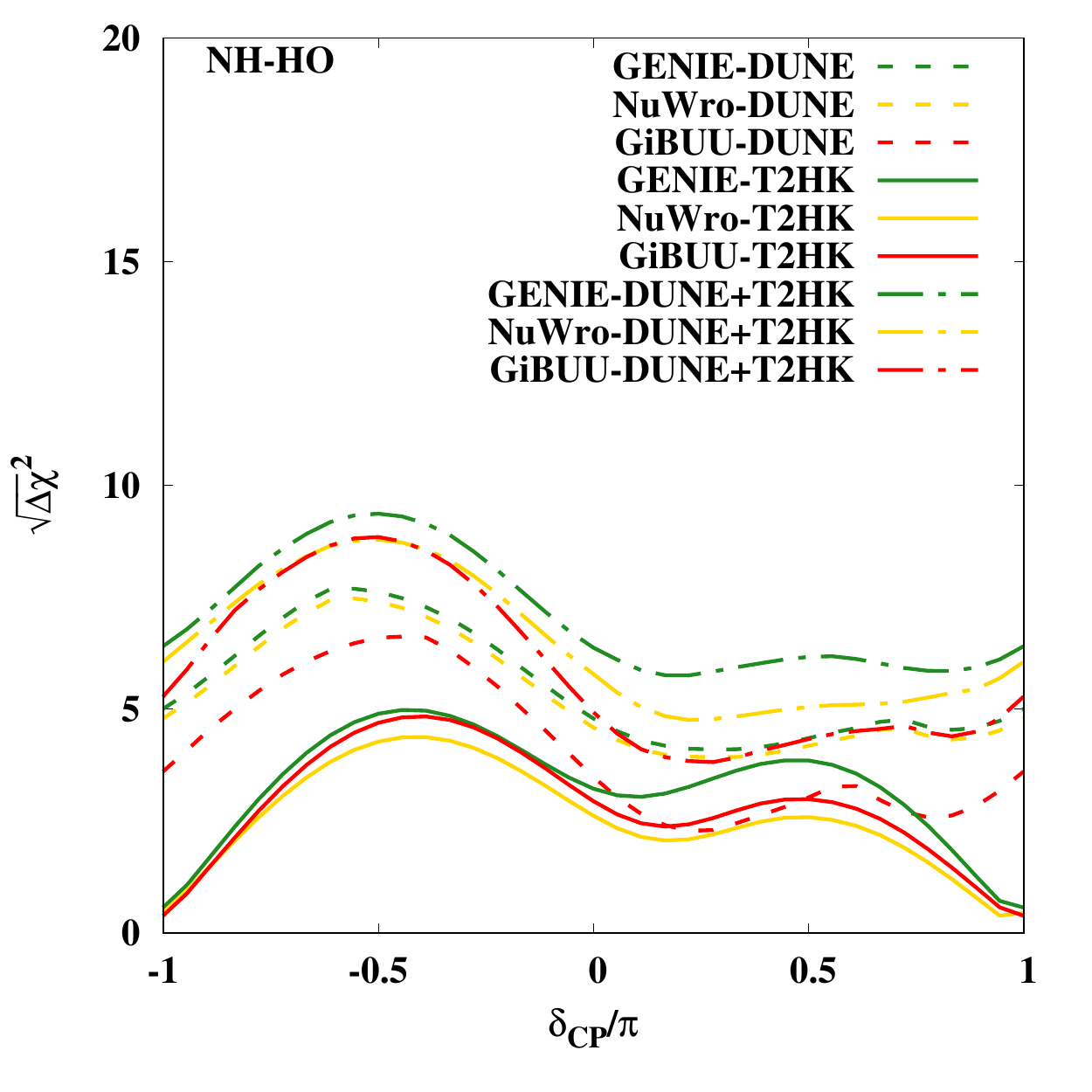} 
		\includegraphics[scale=0.7]{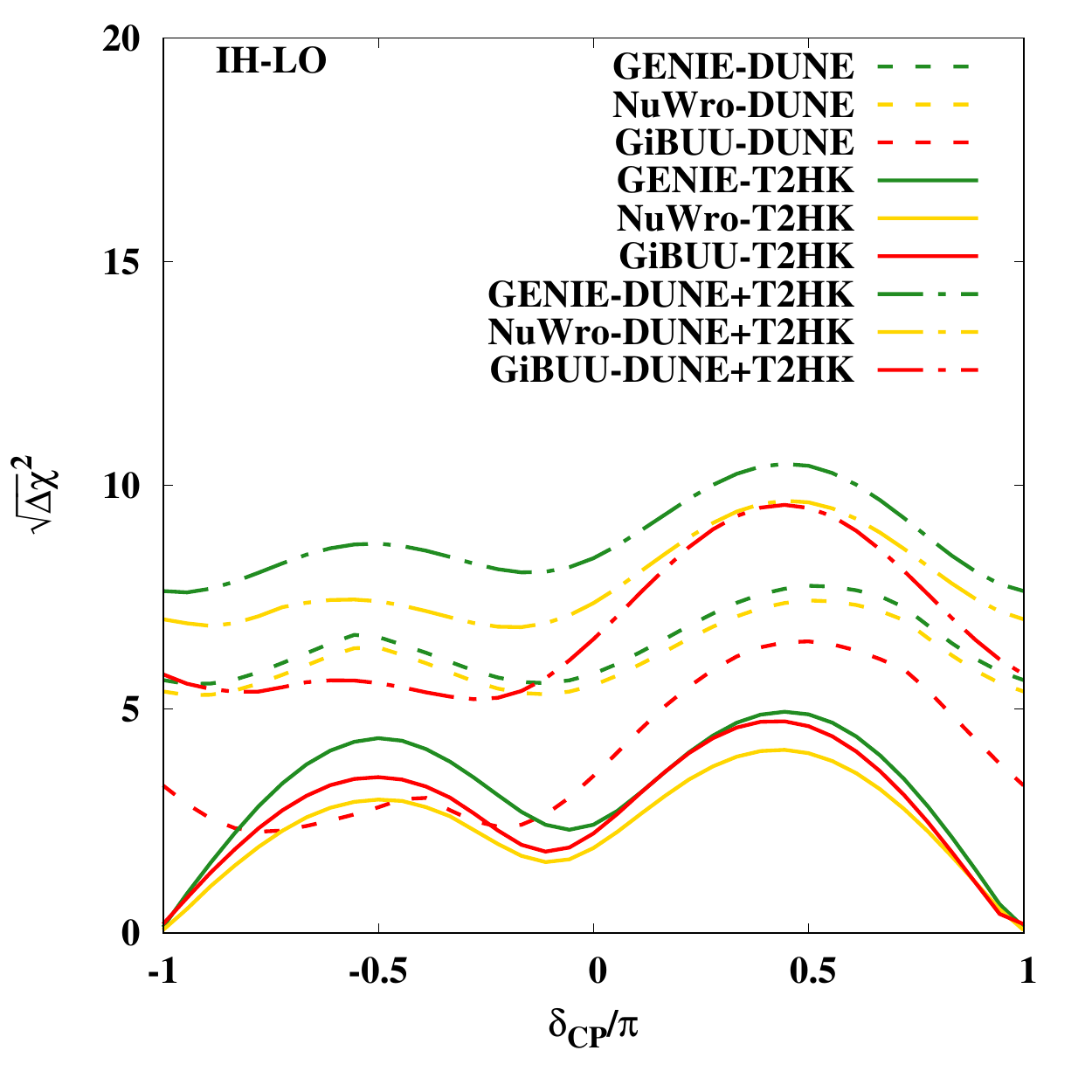} 	\includegraphics[scale=0.7]{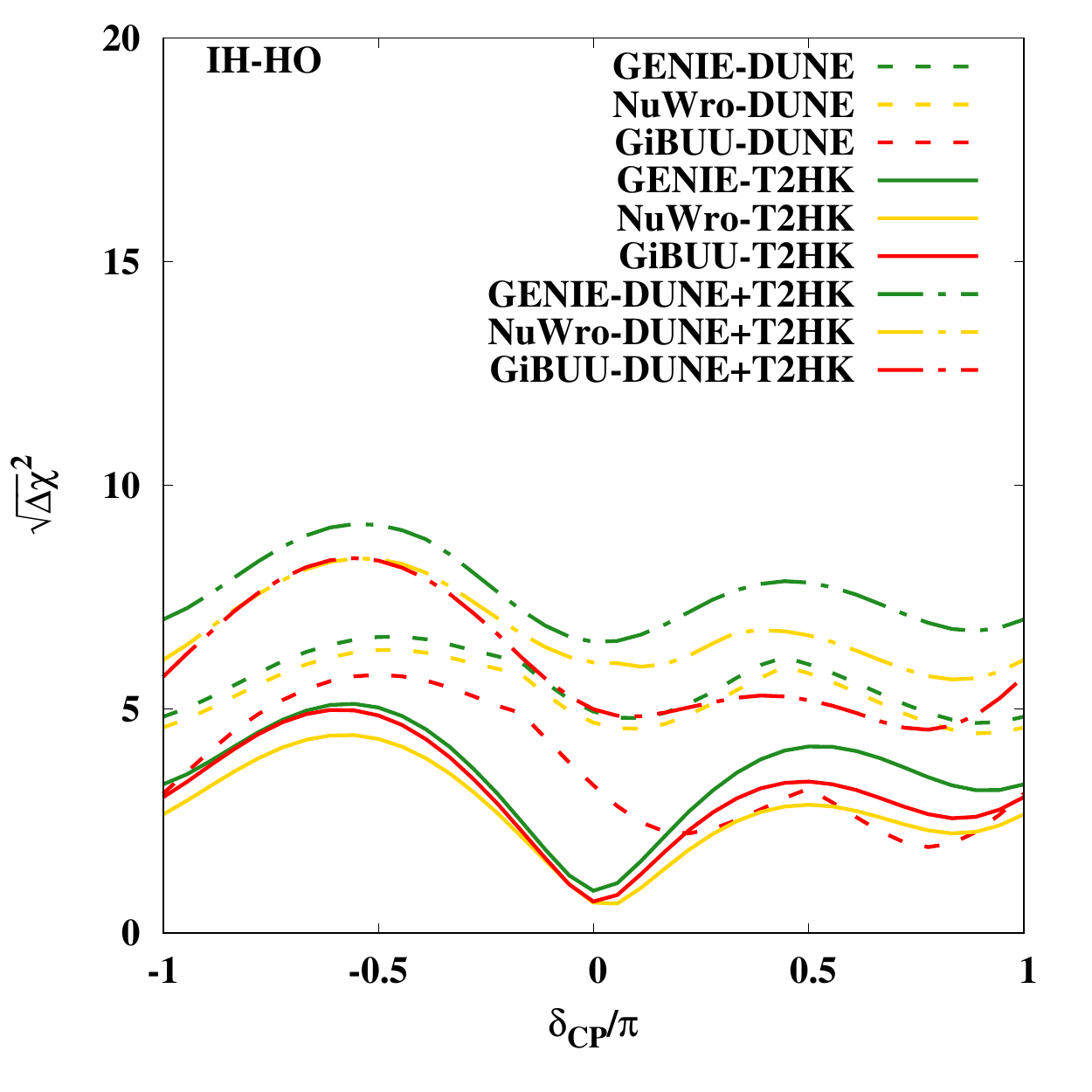} 
		\caption {Octant sensitivity measurement as a function of the true value of $\delta_{CP}$ for NH-LO (top left), NH-HO (top right), IH-LO	(bottom left) and IH-HO (bottom right).}
		\label{fig4}
	\end{figure}

	{In Fig. \ref{fig4}, the top and bottom panels, we show the results for four true hierarchies and octant configurations, namely NH-LO, NH-HO, IH-LO, and IH-HO. 	For the NH-LO scenario (top left panel of Fig. \ref{fig4}), in the case of the DUNE experiment, a difference of around 2$ \sigma $ is observed between GENIE and GiBUU. On the other hand, a variation of more than 1$\sigma$ between GENIE and NuWro at $\delta_{CP} \sim$-0.5/$\pi$. In the combined results of DUNE and T2HK, we observed that there is a difference of around 3$\sigma$ between  GENIE and GiBUU at $\delta_{CP} \sim$-0.5/$\pi$.}

	{For the NH-HO case (top right panel of Fig. \ref{fig4}), we see a small difference in all generators for the T2HK,  but in the case of DUNE, more than 1$ \sigma $  variations between GENIE and GiBUU  at $\delta_{CP} \sim$-0.5/$\pi$. When we combine DUNE and T2HK, we notice around a 2$ \sigma $ difference between GENIE and GiBUU at $\delta_{CP} \sim$0.5/$\pi$.   }

	{For the IH-LO case (bottom left panel of Fig. \ref{fig4}), there is not much of a difference between any of the generators for the T2HK experiment, but for the  DUNE experiment, there is 4$\sigma$ variation between GENIE and GiBUU at $\delta_{CP} \sim$-0.5/$\pi$. The difference between GENIE and GiBUU  is about 3$\sigma$ when DUNE and T2HK are combined. }

	{For the IH-HO case (bottom left panel of Fig. \ref{fig4}), a variation of 1$ \sigma $ between GENIE and NuWro in the case of the T2HK experiment, whereas for the DUNE experiment, we notice a 3$\sigma $  difference between GENIE and GiBUU.
		When DUNE and T2HK are combined, we notice a 2$\sigma $ difference between GENIE and GiBUU at $\delta_{CP} \sim$0.5/$\pi$.}

	\section{Summary and conclusions}
	\label{sec5}
	
	In this paper, we have studied the impact of cross-sectional uncertainty on determining the CP-violation,  mass hierarchy, and  octant degeneracy in the T2HK, DUNE, and DUNE and T2HK using GENIE, NuWro, and GiBUU nuclear models. Because nuclear effects are not completely known, different generators utilize different estimates to account for them, resulting in a variety of results. The physics results' emphasis on the generators' selection will limit the physics analysis targets. Because there is insufficient cross-section data to comprehend the existing nuclear effects, it is necessary to determine cross-sections for diverse nuclei and update the current cross-section data. Every target nuclei' nuclear structure (mass number, atomic number) is unique, posing a significant problem in the correct computation of neutrino-nucleon cross-section.  
	\section*{Acknowledgments} 
	One of the authors, Miss Ritu Devi offers most sincere gratitude to the Council of Scientific and Industrial Research (CSIR), Government of India, for the financial support in the form of Senior Research Fellowship, file no. 09/100(0205)/2018-EMR-I.

\end{document}